\documentclass[10pt,conference]{IEEEtran}
\IEEEoverridecommandlockouts
\usepackage{cite}
\usepackage{amsmath,amssymb,amsfonts}
\usepackage{algorithmic}
\usepackage{graphicx}
\usepackage{textcomp}
\usepackage{xcolor}
\usepackage{listings}
\usepackage{tabularx}
\usepackage{ragged2e}
\def\BibTeX{{\rm B\kern-.05em{\sc i\kern-.025em b}\kern-.08em
    T\kern-.1667em\lower.7ex\hbox{E}\kern-.125emX}}

\newcommand{\fon}[1]{\fontfamily{#1}\selectfont\small	}  
  \usepackage[fixed]{fontawesome5}
\usepackage{diagbox}

\usepackage{booktabs} 
\usepackage{color}
\usepackage{hyperref}
\usepackage{multirow}
\usepackage{pifont}
\usepackage{afterpage}
\usepackage{rotating}
\usepackage{array}
\newcolumntype{x}[1]{>{\raggedright\arraybackslash}p{#1}}

\usepackage{makecell}
\usepackage{graphicx}
\usepackage[font=footnotesize]{caption}
\usepackage[font=footnotesize]{subcaption}
\usepackage[most]{tcolorbox}

\usepackage{multirow}
\usepackage{soul}
\setul{.3ex}{}

\usepackage{url}
\usepackage{lscape}
\usepackage{rotating}
\usepackage{tabularx}

\usepackage{colortbl}
\usepackage{ragged2e}
\usepackage{verbatim}
\usepackage{listings}
\usepackage{arydshln}
\usepackage[neverdecrease]{paralist}
\usepackage{booktabs}
\usepackage{flushend}
\usepackage{enumitem}
\usepackage{CJKutf8}
\usepackage{tikz}
\usepackage{tabulary}

\usepackage{wrapfig}
\usepackage{floatflt}

\usepackage{bigstrut}
\usepackage{threeparttablex}

\usepackage{amssymb}
\usepackage{diagbox}
\usepackage{xcolor}
\usepackage[ruled,vlined, linesnumbered]{algorithm2e}
\definecolor{grayhighlight}{rgb}{.8,0.8,0.8}
\definecolor{LightCyan}{rgb}{0.88,1,1}
\definecolor{LightAmber}{rgb}{0.98,0.89,0.74}
\definecolor{pink}{rgb}{0.9,0,0.9}
\definecolor{gray}{rgb}{0.95,0.95,0.85}
\definecolor{mauve}{rgb}{0.1,0.7,0.2}
\definecolor{bg}{rgb}{0.95,0.95,0.9}
\definecolor{dkgreen}{rgb}{0,0.6,0}
\definecolor{gray2}{rgb}{0.95,0.95,0.95}
\definecolor{gray3}{rgb}{0.20,0.20,0.20}
\definecolor{lightgray}{rgb}{0.6,0.6,0.6}

\definecolor{p1color}{HTML}{8dc7e8}
\definecolor{p2color}{HTML}{9fe4d3}
\definecolor{p3color}{HTML}{ffde59}

\usepackage{colortbl}

\usepackage{balance}
\newcolumntype{L}[1]{>{\raggedright\let\newline\\\arraybackslash}p{#1}} 
\newcolumntype{C}[1]{>{\centering\let\newline\\\arraybackslash}p{#1}} 
\newcolumntype{R}[1]{>{\raggedleft\let\newline\\\arraybackslash}p{#1}}

\definecolor{BlueGreen}{HTML}{117a65}
\newcommand{\revmod}[1]{{\textcolor{black}{#1}}} 

\usepackage[utf8]{inputenc}
\usepackage{epstopdf}

\usepackage{float}
\usepackage{xspace}
\usepackage{tcolorbox}

\newcolumntype{M}[1]{>{\centering\arraybackslash}m{#1}}

\newcommand{\fwname}{\textsc{AutoSimTest}\xspace}
\newcommand{\sagent}{\textit{S-Agent}\xspace}
\newcommand{\magent}{\textit{M-Agent}\xspace}
\newcommand{\eagent}{\textit{Env-Agent}\xspace}
\newcommand{\aagent}{\textit{Analytics-Agent}\xspace}

\newcommand{\etal}{{\textit{et~al.}}\xspace}

\newcommand{\bulletitem}[1]{{\noindent$\bullet$ \textbf{#1}}}

\newtcolorbox[auto counter]{simreq}[1]{title={\bfseries Requirement},drop shadow={black!10!white},
  coltitle=white, colframe=white!25!black, boxsep=2pt,left=4pt,right=4pt,top=3pt,bottom=3pt,  sharpish corners}

  \providecommand\BibTeX{{%
    \normalfont B\kern-0.5em{\scshape i\kern-0.25em b}\kern-0.8em\TeX}}

\newcommand{\citesec}[1]{Section~\ref{sec:#1}}

\lstdefinestyle{jsonlistingstyle}{
  language=json,
  numbers=left,
  stepnumber=1,
  numbersep=8pt,
  tabsize=2,
  showspaces=false,
  basicstyle=\footnotesize\ttfamily,
   numberstyle=\scriptsize,
    showstringspaces=false,
    breaklines=true,
    frame=lines,
    backgroundcolor=\color{background},
}

\newtcolorbox[auto counter]{reqbox}[1]{title={\bfseries #1},drop shadow={black!01!white},
  coltitle=white, 
  leftrule=0.25mm,
  rightrule=0.25mm,
  bottomrule=0.25mm,
  toprule=0.25mm,
  colframe=white!45!black, boxsep=2pt,left=4pt,right=4pt,top=3pt,bottom=3pt,  sharpish corners}

\begin{document}

\title{LLM-Agents Driven Automated Simulation Testing and Analysis of small Uncrewed Aerial Systems}


\author{\IEEEauthorblockN{Venkata Sai Aswath Duvvuru and Bohan Zhang}
\IEEEauthorblockA{\textit{Department of Computer Science} \\
\textit{Saint Louis University}\\
Saint Louis, USA \\
\{venkatasaiaswath.duvvuru, bohan.zhang.1\}@slu.edu}
\and
\IEEEauthorblockN{Michael Vierhauser}
\IEEEauthorblockA{\textit{Department of Computer Science} \\
\textit{University of Innsbruck}\\
Innsbruck, Austria \\
michael.vierhauser@uibk.ac.at}
\and
\IEEEauthorblockN{Ankit Agrawal}
\IEEEauthorblockA{\textit{Department of Computer Science} \\
\textit{Saint Louis University}\\
Saint Louis, USA \\
ankit.agrawal.1@slu.edu}
}
\maketitle
\begin{abstract}
Thorough simulation testing is crucial for validating the correct behavior of small Uncrewed Aerial Systems (sUAS) across multiple scenarios, including adverse weather conditions (such as wind, and fog), diverse settings (hilly terrain, or urban areas), and varying mission profiles (surveillance, tracking). While various sUAS simulation tools exist to support developers, the entire process of creating, executing, and analyzing simulation tests remains a largely manual and cumbersome task. Developers must identify test scenarios, set up the simulation environment, integrate the System under Test (SuT) with simulation tools, formulate mission plans, and collect and analyze results. These labor-intensive tasks limit the ability of developers to conduct exhaustive testing across a wide range of scenarios. To alleviate this problem, in this paper, we propose \fwname, a Large Language Model (LLM)-driven framework, where multiple LLM agents collaborate to support the sUAS simulation testing process. This includes: (1) creating test scenarios that subject the SuT to unique environmental contexts; (2) preparing the simulation environment as per the test scenario; (3) generating diverse sUAS missions for the SuT to execute; and (4) analyzing simulation results and providing an interactive analytics interface. Further, the design of the framework is flexible for creating and testing scenarios for a variety of sUAS use cases, simulation tools, and SuT input requirements.  We evaluated our approach by (a) conducting simulation testing of PX4 and ArduPilot flight-controller-based SuTs, (b) analyzing the performance of each agent, and (c) gathering feedback from sUAS developers. Our findings indicate that \fwname significantly improves the efficiency and scope of the sUAS testing process, allowing for more comprehensive and varied scenario evaluations while reducing the manual effort.
\end{abstract}


\section{Introduction}
Simulation testing is a critical step in the small Uncrewed Aerial Systems (sUAS) development process, to validate the behavior of sUAS across a variety of real-world scenarios \cite{10628470}. This includes, for example, adverse weather conditions (e.g., wind, rain, or fog), diverse terrains (e.g., hilly, flat, urban, and open fields), and varying mission profiles (e.g., long-range, short-range, surveillance, and tracking)~\cite{bondi2018airsim,johnson2002flight,al2019utsim}. Despite the availability of various sUAS simulation tools and software applications~\cite{malik2023carla,anand2021openuav}, simulation testing remains predominantly a manual, or at best partially-automated process. 
As a result, current sUAS simulation testing practices result in three main pain points for sUAS application developers.



First, identifying and designing simulation test scenarios that clearly specify \emph{(a) operational contexts}, including weather conditions, environmental settings, terrain variability, \emph{(b) sUAS sensor specifications} including how noisy each sensor is expected in the scenario, \emph{(c) mission objectives}, including higher-order tasks, sUAS mission modes, and \emph{(d) core safety properties} to test is a challenging and time-consuming task. 

Due to the fact that sUAS operate in complex environments and are expected to perform reliably in unique or even unimaginable real-world scenarios~\cite{erdelj2016uav,scherer2015autonomous}, developers' domain knowledge of sUAS use cases plays an important role. An example, where a software issue caused an incident, is a recent case of a drone crashing in the woods while autonomously tracking a person on an electric scooter. The individual was moving at a moderate speed when suddenly, a dog crossed their path. The drone mistakenly switched its tracking from the person to the dog and crashed into a tree. Video of this drone crash is available on Reddit\footnote{\url{https://www.reddit.com/r/dji/comments/1amf9rg/oops_solid_crash_while_tracking_myself_on_the/}}. 
Such unique scenarios are particularly difficult for sUAS developers to imagine and accurately test during the development phase.

Second, sUAS developers must model each identified scenario to align with the System under Test (SuT) requirements, and configure the simulation tool to simulate environmental context. This process is often manual, repetitive, time-consuming, and prone to human error.

Third, once simulation tests are executed, analyzing flight logs to diagnose abnormal behavior is labor-intensive, requiring a deep understanding of thousands of flight controller parameters \cite{al2022resam}. While tools like PX4 Flight Review~\cite{px4review} plot sensor data automatically, developers still need to interpret the plots and identify issues manually, which requires specific skills and in-depth knowledge of the flight controller. Furthermore, there is no automation framework that enables sUAS developers to analyze simulation artifacts automatically.

\begin{figure*}[t]
    \centering
    \includegraphics[width=0.90\textwidth]{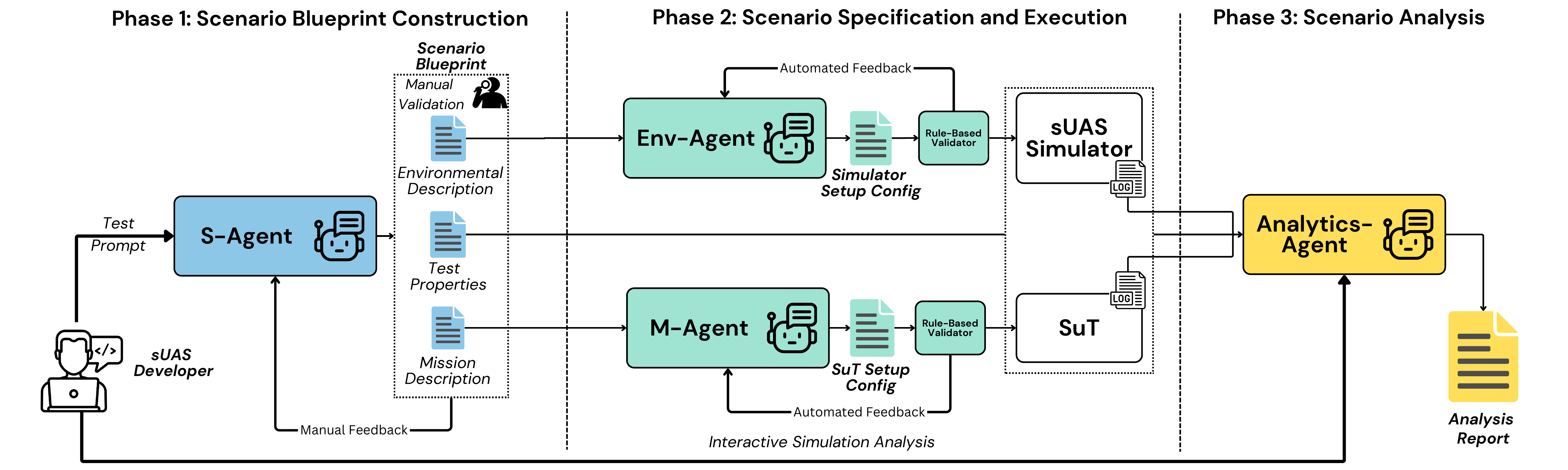}
    \caption{Overview of our \fwname Framework with the 3 main Phases: Scenario Blueprint Construction \revmod{, Manual Validation and Feedback} (blue), Scenario Specification, \revmod{Validation,} and Execution (green), and Scenario Analysis (yellow).
    }
    \label{fig:framework-overview}
    \vspace{-10pt}
    
\end{figure*}

These challenges highlight the need for automated tools and supporting frameworks capable of systematically \emph{specifying}, \emph{generating}, \emph{executing}, and \emph{analyzing} diverse simulation tests to achieve scalability and sufficient simulation test coverage. Therefore, in this paper, we present a novel framework called \fwname as a first step towards automating the simulation testing process for sUAS. We leverage our experience in simulation and field testing within this domain to construct and implement this framework employing recent advancements in Generative AI, more specifically Large Language Models (LLMs) to (a) generate unique test scenarios that incorporate environmental context, mission objectives, and core safety properties to test, (b) model sUAS mission in SuT expected language, (c) configure the simulation environment based on the test scenario's environmental context, and (d) produce a simulation report based on the collected data from scenario execution, as well as provide interactive methods for analyzing simulation results for sUAS developers.

The contributions of this paper are as follows:
\begin{itemize}[leftmargin=*]
    \item \textbf{\fwname}: This is the first work to propose a novel Multi LLM-Agents-based framework to automate the sUAS simulation testing process.   
    \item \textbf{Automated Flight Analysis}: As part of the framework, we introduce a novel \aagent enabling automated and interactive analysis of sUAS flight logs. 
    \item \textbf{Performance and Developers' Perception:} Extensive experiments have demonstrated that \fwname is capable of testing diverse SuT, effectively analyzing flight logs, and providing valuable support to both novice and experienced developers in better interpreting the simulation data.
\end{itemize}

The remainder of the paper is laid out as follows. In \citesec{background} provide a brief introduction to relevant LLM techniques and then, in \citesec{framework}  introduce our \fwname framework. Subsequently, in \citesec{approach} and \citesec{agents} we discuss the different LLM agents part of our framework. We then, in~\citesec{eval}, describe our evaluation setup for addressing feasibility, generalizability, and agent performance, and report on results in \citesec{results}. Finally, we discuss several lessons learned in~\citesec{discussion}, threats to validity in~\citesec{threats} related work in~\citesec{relwork}, and conclude in \citesec{conclusion}.





\section{Background}
\label{sec:background}

This section briefly introduces the LLM techniques used in the design of our \fwname framework (cf.~\citesec{framework}).

\subsection{Retrieval Augmented Generation (RAG)} 
RAG is a method that combines the strengths of pre-trained language models with external knowledge retrieval. Instead of solely relying on the internal knowledge of the language model, RAG retrieves relevant information from external data sources (e.g., a vector database) -- serving as a ``knowledge library'' that the generative AI models can understand. This approach allows the agents to utilize contextually relevant information, especially in specialized domains such as sUAS~\cite{lewis2020retrieval,zhao2024optimizing} to generate text.

RAG consists of four primary steps: Query Generation, Document Retrieval, Context Integration, and Response Generation. First, the input query is processed to generate search queries, which are then used to retrieve relevant documents from an external knowledge base. These documents are integrated back into the language model as additional context. Finally, the model produces a final response that combines its internal knowledge with the retrieved information to generate a more precise and contextually relevant output. 


\subsection{Prompt Engineering} 
Prompt Engineering is a critical technique in the field of Natural Language Processing (NLP) that involves designing and crafting input prompts to elicit desired outputs from language models~\cite{wang2023review,white2023prompt,gu2023systematic,marvin2023prompt,zhou2022large}. The fundamental principle of prompt engineering is to provide clear and context-rich instructions that align with the underlying model's training data. Key strategies to write quality prompts include (a) providing examples to illustrate the request, (b) specifying the desired output format to ensure clarity, and (c) incorporating relevant contextual information to help the LLM understand the prompt and produce the intended output. Prior research in the domain of NLP and Generative AI has shown that the quality of prompt to LLM model greatly influences the quality of generated content~\cite{chen2023unleashing, strobelt2022interactive}.


\section{\fwname - Framework Overview}
\label{sec:framework}

Our \fwname framework supports three main phases of performing sUAS tests. Phase 1, the \emph{Scenario Blueprint Construction}, phase 2, the \emph{Scenario Specification and Execution}, and finally phase 3, the subsequent \emph{Scenario Analysis} and interpretation of results. These phases aim to enhance the sUAS simulation testing process by (a) reducing manual effort in designing and executing scenarios, (b) providing automated analytics support for better understanding of simulation results, while at the same time minimizing reliance on domain-specific knowledge, and finally (c) achieving faster and more frequent simulation testing cycles during development. 

Figure~\ref{fig:framework-overview} provides an overview of the three phases of the framework. Each phase consists of specialized LLM-based AI agents to automate parts of the testing process. All agents share the generated data to provide the necessary automation support. In the following, we provide a brief overview of the main functionality of each phase and discuss the role of each AI agent, before we provide further design and implementation details in~\citesec{approach} and \citesec{agents}.

\subsection{Phase 1 -- Scenario Blueprint Construction}
\label{subsec:phase1}
As a starting point, the sUAS developer provides the high-level objectives of the simulation tests that should be created in natural language, as textual input. Depending on the focus of the simulations to be performed, this input can range from very specific goals, such as \emph{``testing the computer vision model of the system under foggy weather conditions''}, to more general objectives, such as \emph{``evaluating the navigation capabilities of the system''}. This crucial input then serves as a part of the prompt for the first agent, the Scenario-Gen-AI-Agent (\sagent) which is tasked with generating a scenario blueprint based on its knowledge of past sUAS incidents in the real world. The \sagent utilizes a Retrieval-Augmented Generation approach to consider real-world sUAS incidents to generate relevant scenario blueprints (cf.~\citesec{approach}). 

The output of the \sagent includes the textual specifications of the environment, the mission that the SuT must execute, and importantly, test properties to evaluate the SuT. 
For instance, a developer might provide input to the \sagent as \emph{``Test the ability of a drone to track a missing person during a search-and-rescue mission''}. In response, the \sagent generates:

\begin{tcolorbox}[colback=blue!4,colframe=blue!30,fontupper=\fon{cmss},boxsep=2pt,left=6pt,right=6pt,top=5pt,bottom=0pt]

\begin{itemize}[leftmargin=*]
\addtolength{\itemindent}{0.2em}
\item[{\small\faComment*[regular]}]\textbf{
 Environmental Description}: A densely forested area with tall trees, uneven terrain, and potential obstacles like fallen logs, branches, and wildlife activity, alongside dynamic conditions such as varying light levels, fog, and light rain.
\vspace{0.35em}
\item[{\footnotesize\faComment*[regular]}]\textbf{sUAS Mission Description}: The SuT should be tasked with locating a missing hiker in a forest environment, navigating through waypoints including search patterns, obstacle avoidance maneuvers, and target identification. 
\vspace{0.35em}
\item[{\footnotesize\faComment*[regular]}]\textbf{Test Properties}: Specific metrics, such as target detection and identification accuracy, obstacle avoidance efficiency, flight stability in varying weather, sensor performance under different lighting, and overall mission completion time.
\end{itemize}
\end{tcolorbox}

\revmod{Before being fed into the second phase of the pipeline, a stage-gate is added where the scenario blueprints undergo manual inspection and validation by a human, for example, developers, testers, or safety engineers. This ensures that the generated scenarios are accurate and do align with testing needs and respective safety-related requirements (e.g., operational boundaries of sUAS related to environmental conditions). Developers can provide feedback to \sagent to refine and adjust specific components of the blueprint such as \emph{``increase the mission complexity''} or \emph{``add lighting variability to the environment''}. This feedback process ensures early identification and correction of scenarios that do not meet developers' objectives. }

\subsection{Phase 2 -- Scenario Executable Script Generation}
This phase focuses on automatically generating the scenario execution scripts as per the sUAS developer's testing infrastructure. This phase consists of two specialized agents each taking over and creating one crucial part required for scenario execution: the SuT Mission Execution Script, and Simulation Tool Configuration Script


\textit{\eagent}: Simulation tools such as AirSim~\cite{shah2018airsim}, Gazebo~\cite{gazebo}, and DroneReqValidator~\cite{agrawal2023requirements,zhang2023dronereqvalidator} require configuration scripts, typically in JSON or XML format, to initialize the simulation environment. These scripts define various parameters necessary for the simulation, such as weather conditions, the origin of the simulation environment, drone characteristics (e.g., Quad-copter or fixed-wing), and their home or starting geolocations. 
Therefore, the \eagent takes the high-level textual environment description from phase 1, provided by the \sagent, as input and utilizes it as a primary prompt to generate a configuration script. 

This generated output is used to initialize the 3D environment of the simulation tool according to the scenario execution requirements. For instance, if a scenario needs to be executed in foggy weather, or an urban environment, the \eagent will generate the necessary tool configuration to simulate these conditions. However, the fidelity of the weather conditions and the realism of the environment, e.g., buildings, depends on the fidelity of the simulation tool itself. 
Figure~\ref{fig:eagent} shows a sample script generated by the agent to initialize the AirSim simulation tool with weather conditions and a vehicle. 

\begin{figure}
    \centering
    \begin{minipage}{.40\columnwidth}
        \centering
        
        \input{listings/listing-EnvGenerator-Airsim}
        \caption{\eagent output}
        \label{fig:eagent}
    \end{minipage}
    \hfill
    \begin{minipage}{.55\columnwidth}
        \centering
        
        \input{listings/listing-MissionGenerator}
        \caption{\magent output}
        \label{fig:magent}
    \end{minipage}%
    
    \vspace{-2em}
\end{figure}

\textit{\magent}: Missions define the tasks and actions that sUAS must perform, serving as inputs to the SuT. These missions specify tasks like searching, tracking, or flying in patterns to capture imagery. Typically, sUAS developers use open-source tools such as QGroundControl~\cite{qground} or MissionPlanner~\cite{mplanner} for creating these missions. These tools allow users to specify waypoints, set altitudes, define actions at each waypoint (e.g., taking photos or adjusting speed), and incorporate safety protocols, such as return-to-home triggers, or no-fly zones. However, the manual process is time-consuming and prone to human error due to the vast number of settings and configurations required. Moreover, the complexity of these settings increases with the increasing complexity of the SuT.
Therefore, similar to the \eagent, the \magent takes the high-level textual mission description from the \sagent and transforms it into specific, executable missions for the SuT.  Figure \ref{fig:magent} shows a sample output to execute a PX4 mission. 

\revmod{\textit{Validation and Error Correction}: Given the susceptibility of current LLMs to issues such as hallucinations~\cite{ji2023survey}, it is important to validate the generated scripts.
For this purpose, we introduce an automated validation step, before scripts are executed, using a set of pre-defined rules. Developers can create these rules based on their sUAS mission and simulation tool requirements. For instance, rules can be designed to detect basic syntactical errors in the generated scripts, such as incorrect formatting, missing fields, or invalid/out-of-range parameter values in mission. When issues are detected, the validator provides immediate feedback to the respective agent, prompting it to regenerate a corrected script with specific error details. This feedback loop ensures that any errors introduced by the LLMs are resolved within the framework and do not propagate to subsequent stages.}

Finally, after the mission is validated for execution and the simulation environment is initialized with environmental context as per the scenario, either a developer can manually trigger the simulation execution in SuT or it can also be automated using automation scripts. 

\subsection{Phase 3 -- Scenario Analysis Generation} 
The execution of a simulation scenario typically generates simulation results in the form of SuT logs, which developers record in their code base for debugging purposes. These SuT logs contain high-level explanatory messages, such as \texttt{\footnotesize Mission Dispatched Successfully} and \texttt{\footnotesize Moving to the next waypoint}, as well as low-level log messages that include event names and their timestamps. These logs are crucial for developers to understand and debug any issues related to SuT execution as per the system requirements. In addition to SuT logs, the underlying flight controller used by developers, such as PX4~\cite{px4} or ArduPilot~\cite{ardupilot}, generate their own logs called flight logs. These flight logs contain time-series sensor data, and warnings or messages generated by flight controllers. This data is crucial for conducting lower-level analysis and understanding the behavior of the vehicle. Developers analyze these logs to interpret the simulation results. Analyzing time-series data from hundreds of flight controller parameters, however, requires extensive domain expertise and knowledge. Therefore, the primary objectives of \aagent are twofold:
\begin{itemize}[leftmargin=*]
\item \textit{Automated Scenario Analysis}: Generate scenario analysis reports from data produced as part of the scenario execution.
\item \textit{Bridge Developer's Knowledge Gap in Analysis}: Provide developers with interactive methods to explore and understand complex simulation logs. This interactive analysis aims to bridge the knowledge gap by allowing developers to ask high-level analytics questions and receive automated analysis from \aagent in the form of a small set of relevant parameters to investigate, out of hundreds of parameters.

\end{itemize}


\section{Agents' Knowledge-Base and Prompts}
\label{sec:approach}

In this section, we discuss two main components of our framework: the knowledge base and custom prompt design.
\subsection{Knowledge Base}

\textit{\sagent knowledge base}: To automatically generate scenarios that aim to detect specific vulnerabilities in the SuT, especially the ones that are common in the real world, we created a robust database drawing from publicly available information on sUAS real-world incidents. The sources we used to build the S-Agent's knowledge base are listed in Table~\ref{table:s-agent-knowledge-base}. We used web scrapers to collect the data from sources and semi-autonomously cleaned them to create the knowledge base. This knowledge base provides the agent with realistic scenarios and contexts that have previously caused issues in sUAS operations. This data-driven approach enables the \sagent to generate simulation testing scenario blueprints that are informed by actual events and historical data. This data-driven approach is also extensible to other domains, e.g., autonomous vehicles (further discussed in Section \ref{sec:discussion}).

\begin{table}[htbp]
    \centering
    \begin{tabular}{|c|p{4.45cm}|r|r|}
\hline
\textbf{No.} & \textbf{Incident Source} & \textbf{Incident \#} & \textbf{Token \#} \\
\hline
1 & UK Air Accident Investigation Rep.~\cite{UASAirAc93:online} &  81 & 6280 \\
\hline
2 & Wikipedia List of UAV incidents \cite{Listofun27:online} & 74 & 1203 \\
\hline
3 & NASA ASRS Report \cite{asrsarcn42:online} & 50 & 11316 \\
\hline
\end{tabular}
    \caption{Scenario Incident Sources to build \sagent Knowledge Base.}
    \label{table:s-agent-knowledge-base}
    \vspace{-1.5em}
\end{table}

\textit{\aagent knowledge base}: With extensive simulation data and hundreds of parameters in each log, the \aagent must select the relevant parameters and log sections to generate relevant responses. This requires the agent to have domain knowledge to interpret the log data and understand its content and meaning.

To build this second knowledge base, we developed a dataset that encodes the meaning and interpretation of hundreds of parameters from the two most popular open-source flight controllers: Px4 and ArduPilot. The Px4 flight controller codebase\footnote{\url{https://github.com/PX4/PX4-Autopilot/tree/main/msg}} provides a comprehensive description of each logged parameter, with detailed comments explaining the meaning and interpretation of each parameter. We used a Python script to parse the data structure of each message, which contains the parameter name and a brief description as comments. This domain-specific information enables the \aagent to leverage RAG techniques to identify and analyze contextually relevant parameters within the flight logs based on user queries and scenario test properties.

\begin{table*}[h]
\centering
\renewcommand{\arraystretch}{1.12}

\begin{tabularx}
{\textwidth}{|L{0.95cm}|L{2.7cm}|L{2.90cm}|L{3.2cm}|L{2.65cm}|L{3.1cm}|}
\hline
\diagbox[width=4.9em,height=3em]{\textbf{Part}}{\textbf{Agent}}
& 
\textbf{S-Agent} & 
\textbf{M-Agent} & 
\textbf{Env-Agent} & 
\textbf{A-Agent (Automated)} & 
\textbf{A-Agent (Interactive)} \\ \hline \multirow{5}{1.5cm}{\textbf{Agent}\\\textbf{Goals}} & Act as a Test Engineer to generate scenario blueprints that include 
\begin{itemize}[leftmargin=*]
    \item[-] Environment
    \item[-] Mission
    \item[-] Test Props\vspace*{-\baselineskip}
\end{itemize} & Act as an Automation Engineer to generate a SuT mission script based on scenario blueprint & Act as an Automation Engineer to generate sim tool script based on scenario blueprint &  Act as a Data Analyst to analyze simulation log data and explain how the test properties are affected &\multirow{5}{1.5cm}{N/A}\\\hline
\multirow{3}{1.5cm}{\textbf{User}\\\textbf{Goals}} &\textit{Example}: Generate a search-and-rescue mission scenarios  & \multirow{3}{1.5cm}{N/A} & \multirow{3}{1.5cm}{N/A} & \multirow{3}{1.5cm}{N/A} & \textit{Example}: Explain all px4 parameters that might have an impact of high wind \\ \hline
\multirow{2}{1.5cm}{\textbf{Sample}} & cf. Example blueprint in Section \ref{subsec:phase1} & cf. Figure~\ref{fig:magent} & cf. Figure~\ref{fig:eagent} & \multirow{2}{1.5cm}{N/A} & \multirow{2}{1.5cm}{N/A}\\\hline
\multirow{6}{1.5cm}{\textbf{Rules}} & 
\begin{itemize}[leftmargin=*]
    \item[-] Blueprint Completeness
 \end{itemize}
& 
\begin{itemize}[leftmargin=*]
    \item[-] Script Format Validity
    \item[-] Valid Geo-loc.
    \item[-] Valid Waypoints 
    \item[-] Velocity = [0,30] mph
    \item[-] Altitude $\leq 400 ft$\vspace*{-\baselineskip}
    \end{itemize}
& 
\begin{itemize}[leftmargin=*]
    \item[-]  Script Format Validity
    \item[-]  Wind = [0,50] mph
    \item[-]  Light Intensity = (0,10)
    \end{itemize}

& 
\begin{itemize}[leftmargin=*]
    \item[-] Analysis Completeness 
\end{itemize}

& 
\begin{itemize}[leftmargin=*]
    \item[-] Analysis Completeness 
\end{itemize}\\ \hline

\multirow{3}{1.5cm}{\textbf{Ext. Data}} &  List of past real world sUAS incidents & \multirow{2}{1.5cm}{N/A} & \multirow{2}{1.5cm}{N/A} & \multicolumn{2}{|c|}{Flight Controller Codebase and Documentation} \\ \hline
\end{tabularx}
\caption{Overview of the prompt design for four agents part of \fwname.}
\label{tab:promptdesign}
\vspace{-2em}
\end{table*}

\subsection{Prompt Design}

We use a pattern-based approach~\cite{white2023prompt} to design prompts for each agent. Our prompt design for agents in the framework consists of four essential patterns:
\begin{itemize}[leftmargin=*]
\item \textit{Agent's Goals}: We utilize the \emph{Persona Pattern} \cite{olea2024evaluating} to instruct the agents about their primary goals when generating output. For instance, we use the ``Act as a Simulation Testing Engineer'' persona to design sUAS scenarios, or ``Act as a Data Analyst'' to analyze the simulation logs. The Persona Pattern helps the agent to identify the details it should focus on when generating the output.

\item \textit{Sample of Expected Output}: We utilize the \emph{Template Pattern} to instruct the agent about the format of the expected response. The \sagent, \magent, and \eagent include this component, especially because the output is directly fed into another sub-system and must match the sub-system's input format for automated execution.

\item \textit{User Goals}: 
We utilize the \emph{Context Manager} Pattern to instruct the LLM about the User Goals to fulfill when generating the output. The prompt includes instructions from the user's perspective to tailor the output, such as generating a scenario blueprint for a complex search-and-rescue mission or a simpler package-delivery mission. The \sagent and \aagent include this component in their prompt since the user initiates the process using \sagent and consumes/interacts with the output of the framework using the \aagent. Further, these agents utilize RAG to include the contextual information from their respective knowledge base, based on user goals, to augment targeted and relevant information.  

\item \textit{Rules}: Additionally, we utilize the \emph{Error Identification Pattern} to force the LLM to validate the generated output against a set of predefined rules or conditions. For instance, the sUAS mission script should not allow sUAS to fly over 400ft (safe altitude). This validation process ensures that the generated output meets the required criteria and maintains consistency in the LLM output.

\end{itemize}

\vspace{0.25em}
Table \ref{tab:promptdesign} summarizes the prompt design for each agent, including the four components and the data utilized from their knowledge base to generate a response. The following sections provide details about the design and architecture of each agent.

\section{Agents' Design}
 \label{sec:agents}
\subsection{\sagent}

First, the agent uses an embedding model \cite{lee2024nv} to vectorize the knowledge documents (cf.  Table \ref{table:s-agent-knowledge-base}) and store them in a vector database (\textit{RAG Knowledge Store}) to perform search queries on it. 
 Second, when an sUAS developer provides input to the agent to generate a scenario blueprint as per their testing needs, the agent vectorizes the textual input (\textit{Prompt Vector}) and performs a cosine similarity search in the \textit{RAG Knowledge Store} to identify a set of real-world sUAS incidents relevant to the user's testing context. For example, if a user wishes to test a search-and-rescue operation, the agent will find the details of accidents during search-and-rescue operations and utilize that to formulate User Goals. In addition to the \textit{User Goals}, the agent incorporates other components (\textit{Agent's Goals}, \textit{Sample of Expected Output}, and \textit{Rules} as described in Table \ref{tab:promptdesign}) of the prompt to form a comprehensive prompt and sends it as input to the LLM to generate a scenario blueprint.
ection{\magent and \eagent}

The \magent and \eagent agents utilize specific parts of the scenario blueprint as their goals to generate scripts. The \magent uses the Mission Description section as its goals, while the \eagent uses the Environment section. Since these agents act as translators, converting textual descriptions into a language that the SuT and simulation tool can interpret, they don't require \textit{User Goals}. However, to ensure their generated output is compatible with the SuT and simulation tool, they include \textit{Sample of Expected Output} and \textit{Rules} to follow when generating output. The detailed breakdown of the prompt design is shown in Table~\ref{tab:promptdesign}.


\subsection{\aagent}
\begin{figure}[t!]
    \centering
    \includegraphics[width=0.9\columnwidth]{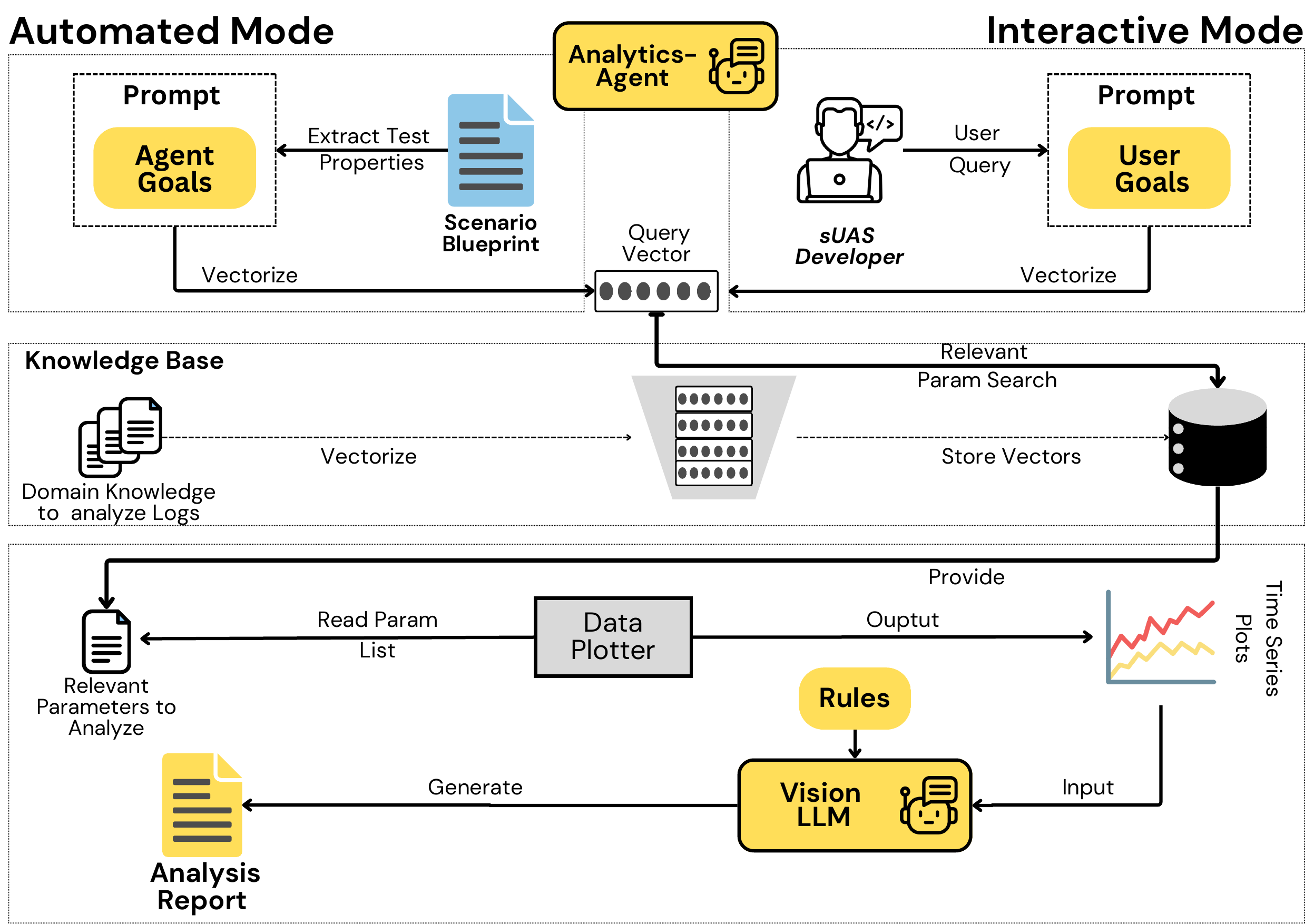}
    \caption{Schematic overview of the main parts of the \aagent.}
    \label{fig:a-agent}
    \vspace{-15pt}
\end{figure}

The design of the \aagent is divided into two main, logically separated components, as shown in Figure \ref{fig:a-agent}. The \emph{Automated Mode} component is responsible for analyzing simulation logs based on the test properties generated by \sagent and producing a comprehensive scenario analysis report without direct interaction with the user. Complementing this first part, the second part of the \aagent, the \emph{Interactive Mode} component allows sUAS developers to start new analyses that may not be part of \emph{Automated Mode} analysis. 

Both components utilize a shared knowledge base to identify the essential simulation log data to analyze. 
The primary distinction between the two modes lies in the formulation of prompts. In Automated mode, the agent uses \textit{Agent's Goals} and \textit{Rules}, whereas Interactive mode relies on user input, designated as \textit{User Goals} and \textit{Rules}, in the prompt. Based on the Analysis Goals, whether from the User or the Agent, the agent performs a semantic search within the knowledge base to identify flight controller parameters to analyze such as velocity, pitch angles, GPS position, and altitude.

Upon identifying the core flight parameters, the agent initiates an automated script to plot these parameters, with the x-axis representing the timestamp and the y-axis representing the parameter values. These plots, along with the prompt, are then provided as input to the Vision LLM model to analyze the plot images and generate an analysis report. \aagent saves this report for developers' debugging and analysis purposes.



\section{Evaluation}
\label{sec:eval}
We used our \fwname framework to investigate the following research questions: \vspace{-0.35em} \\


   \noindent\emph{ -- RQ1:  To what extent is our framework applicable for conducting real-world sUAS simulation testing, and how well can it be applied across diverse SuT that use different flight controllers and simulation tools for testing?}

    To answer this research question, we conduct experiments to first evaluate (a) the feasibility of \fwname to test different SuT that utilize widely-used flight controllers such as PX4~\cite{px4} and ArduPilot~\cite{ardupilot}, and second (b) its ability handle diverse common sUAS use cases (ref Table \ref{tab:use-cases}).

    \vspace{0.4em}
  \noindent\emph{-- RQ2: To what extent can our framework with RAG be utilized for generating relevant test scenarios blueprints across sUAS use cases and investigating sUAS failures?}

We conducted a second set of experiments to quantitatively assess the performance and effectiveness of our RAG approach, for both the \sagent and the \aagent. We used the standard Retrieval-Augmented Generation Assessment (RAGAs) framework \cite{es2023ragas}, a reference-free RAG pipeline evaluation method that utilizes a separate LLM as a critic to judge the responses generated by our agents. 
We collect the following metrics.

\begin{itemize}[leftmargin=*]
    \item \emph{Context Precision and Recall}:  These metrics measure the effectiveness of  RAG in retrieving relevant information from the knowledge base. \emph{Context Precision} evaluates the proportion of retrieved information that is relevant, while \emph{Context Recall} evaluates the proportion of relevant information.
    \item \emph{Response Faithfulness}: This metric assesses whether an AI-agent generate outputs that are grounded in the information gathered from their knowledge bases. 
    \item \emph{Response Relevancy}: This metric evaluates the completeness of the final responses generated by an agent. 
\end{itemize}
To compute Response Relevancy and Faithfulness, we employ the Llama 3 LLM~\cite{llama3}, a 3 trillion parameter model, as a critic to evaluate responses of \sagent and \aagent. Further, we also  investigate the quality and correctness of the \aagent output using flight logs as described in \citesec{setup}. 




    
    

\vspace{0.4em}
 \noindent\emph{-- RQ3: How do developers perceive the generated scenarios, and does the interactive analysis support developers in better understanding the simulation results?}

To address this research question, we conducted interviews (each approx. lasting 1 hour) with sUAS developers to understand current challenges in sUAS simulation testing and their perception about \fwname and especially how the interactive analysis can support them during simulation log analysis. We deployed our entire framework on the cloud and asked participants to especially interact with \revmod{\sagent and} \aagent. Participants provided feedback and opinions on the framework's usability in sUAS simulation testing.

Next, we discuss our strategy, tools, and techniques we utilized to implement the framework that we utilize for conducting experiments.

\subsection{Evaluation Setup}
\label{sec:setup}
\textbf{\fwname Framework Implementation:}
\label{sec:implementation} We leverage the advanced capabilities of the open-source Phi-3 family of LLMs~\cite{abdin2024phi} designed by Microsoft to implement our framework. Specifically, we utilize the Phi-3-medium-128k-instruct 
LLM~\cite{microsof_phi3_medium:online}, a model with 14 billion parameters. 
\begin{figure}[b!]
\vspace{-1em}
    \centering
    
    \frame{\includegraphics[width=0.80\columnwidth]{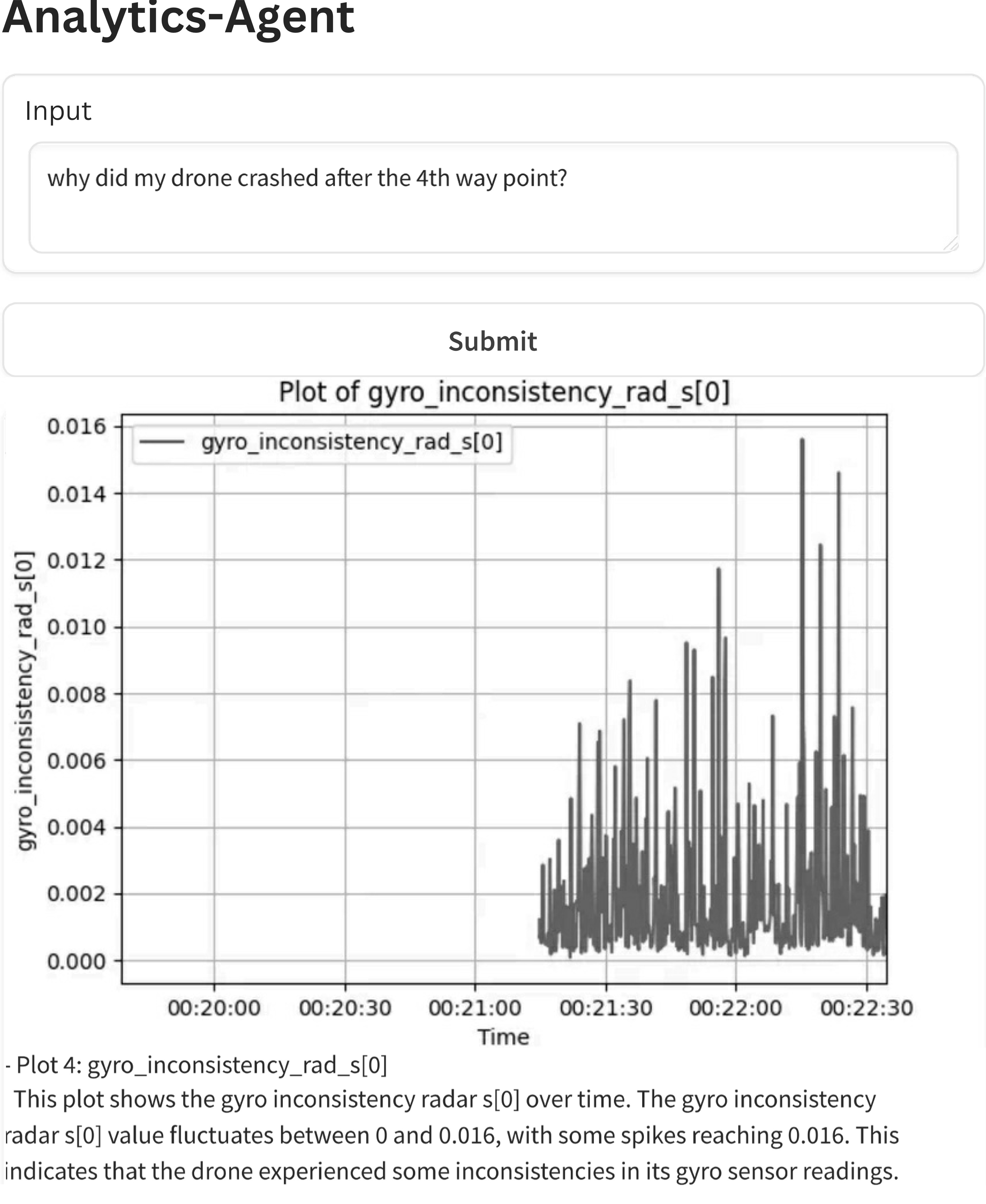}}
    \caption{User Interface for developers to interact with \aagent. Top: User Query; Middle: Generated Plot; Bottom: Generated Analysis Report}
    \label{fig:ui}
    \vspace{-1em}
\end{figure}
In order to support analysis of image data, we use the Phi-3-vision-128k-instruct model \cite{microsof_phi3Vision__medium:online}, which integrates an image encoder, connector, projector, and a Phi-3 Mini language model, all within 4.2 billion parameters. For embedding and similarity matching, we use sentence-transformers/all-mpnet-base-v2 model \cite{reimers-2019-sentence-bert} and FAISS~\cite{johnson2019billion} for vector database management.  We deployed our framework on Gradio.App~\cite{Gradio2:online}, a cloud platform for ML models. We use this deployment in our perception study to gather feedback from sUAS developers. Figure \ref{fig:ui} shows the interface our participants used in the study.

\textbf{SuT:} We designed two sUAS, SuT-Px4 and SuT-Ardu, based on Px4 and Ardupilot flight controllers, respectively. These flight controllers were chosen due to their widespread use, comprehensive documentation, and open-source license. For simulation, we employed AirSim (compatible with Px4) and Ardupilot-SITL (compatible with Ardupilot). We developed our SuT for city surveillance use case with two features: 
\begin{itemize}[leftmargin=*]
    \item Way-point-based Navigation - SuT follows predefined way-points to surveil an area.
    \item Autonomous Navigation - SuT determine its grid-based flight path based on the shape of surveillance area.
\end{itemize}

\textbf{sUAS Use Cases:} e examined the effectiveness of \fwname across popular sUAS use cases, as summarized in Table \ref{tab:use-cases}. These specific use cases were selected due to their popularity in both academic research and industrial products.

\begin{table}[htbp]
    \centering
    \begin{tabular}{|L{4cm}|L{3.8cm}|}
    \hline
         \textbf{sUAS Use Case} & \textbf{References}  \\ \hline
         UC1: Search-and-Rescue & \cite{search1, search2, search3}\\ \hline
         UC2: Precision Agriculture & \cite{agri1,agri2,agri3}\\ \hline
         UC3: Env. Monitoring &  \cite{env1,env2,env3}\\ \hline
        UC4: Surveillance & \cite{surv1,surv2,surv3} \\ \hline
         UC5: Package Delivery & \cite{pack1,pack2,pack3}\\ \hline
    \end{tabular}
    \caption{Common sUAS use cases with references in the literature}
    \label{tab:use-cases}
\end{table}

\textbf{SuT Flight Logs: } We created a dataset of 7 PX4 flight logs with artificially injected sensor failures in our PX4-SuT. \revmod{We used PX4 utilities to inject these errors \cite{SystemFa82:online}, as they are widely used during the simulation testing phase to model common real-world sensor failures, such as sudden GPS loss}. WEach flight log containsone of sensor failure including GPS, Accelerometer, Gyro, Magnetometer, Battery, Barometer, and AirSpeed. We use these flight logs to evaluate the ability of \aagent in detecting sensor failures (Section \ref{sec:agent-qualitative-analysis}).

\section{Results and Analysis}
\label{sec:results}
\subsection{RQ1 - Applicability and Generalizability}
\label{sec:evRQ1}
\subsubsection{\textbf{Applicability}}
 To demonstrate the general applicability of \fwname for sUAS simulation testing, we apply each phase to our two SuT, for a common use case and analyze the resulting output. Starting in phase 1, we used the \sagent to design a scenario blueprint for a city surveillance use case. We used a very simple ``Generate a city surveillance'' scenario as \emph{User Goals} of the prompt. This blueprint serves as input for phase 2 to generate executable scripts for our SuT, specifically SuT-PX4, and SuT-Ardu, using the mission agent, and for simulation tools, specifically AirSim and ArduPilot-SITL, using the environment agent. We provide the \emph{Sample of expected outputs} to these agents to generate valid executable scripts. The generated scripts were used to automatically execute SuT-PX4 missions in AirSim~\cite{shah2018airsim} and SuT-Ardu missions ArduPilot-SITL~\cite{ardusitl}. 
 
\textit{Results and Analysis}: The scenario blueprint, presented in Table~\ref{tab:surviellance-blueprint}, outlines the environment, sUAS mission, and test properties. The executable scripts generated by the \magent and \eagent were validated, and confirmed to contain accurate geolocation waypoints for New York City (Latitude: 40.7128, Longitude: -74.0060) and wind speed settings of 15m/s in the AirSim simulation script (see supplementary materials for complete scripts). We were able to use generated mission scripts in both simulation tools without making any changes and collected the flight logs for analysis. This initial set of tests demonstrates the capability of our framework to generate executable scenarios in simulation. In the next step, we discuss how the framework can be applied to test SuT in an extended more broader manner.


\begin{table}[t]
    \centering
    
    \begin{tabular}{|c|p{6cm}|}
    \hline
    \rowcolor{lightgray}
    \textbf{Category} & \textbf{Description} \\
    \hline
     Environment & 
    \begin{itemize}[nosep,leftmargin=*]
        \item \textit{Location}: New York City
        \item \textit{Weather}: Wind = 15m/s
        \item \textit{GPS Quality}: High
        \item \textit{Obstacles}: Buildings, PowerLines
    \end{itemize} \\
    
    \hline
     Mission & Monitor traffic patterns, and assist law enforcement in maintaining public safety. \\
    \hline
    Test Property & Flight Stability in Wind \\
    \hline
    \end{tabular}
    \caption{Scenario Blueprint for City Surveillance Use Case}
    \label{tab:surviellance-blueprint}
    \vspace{-1.5em}
\end{table}

\subsubsection{\textbf{Generalizability across sUAS Use Cases}}
\revmod{As a first step, to demonstrate applicability, we applied our framework to 5 different sUAS use cases, each of which requiring different properties to be tested and simulations to be configured accordingly (cf. Table \ref{tab:use-cases}). For each use case, we used the \sagent to generate 5 scenario blueprints using a simple prompt such as \emph{``Create a scenario to test my sUAS in delivering emergency supplies during a natural disaster.''} To validate our framework's ability to generate correct executable mission and simulation scripts, we adopted a model-based approach using a JSON-schema to define valid scripts for hypothetical SuTs in each use case. This choice aligns with the widespread use of structured JSON messages as a standard for mission inputs in both commercial \cite{Dronelin76:online} and academic sUAS systems \cite{cleland2020requirements,al2023configuring}.
}

\revmod{When generating mission scripts where we} leveraged this model to specify the \emph{Sample of Expected Output} component of the prompt for the \magent. To further enhance the prompt, we introduced new rules that capture specific requirements for each use case. For instance, we added a rule that ``Missing Person'' should be a valid geolocation when generating mission scripts for search-and-rescue missions. Similarly, we included a rule that the distance to the delivery location should be fewer than 2 miles from the home location of a drone when generating missions for package delivery. We followed a similar approach to generate scripts using \eagent for AirSim simulation tool settings. \revmod{As a result, using our framework, we created a total of 25 Scenario Blueprints, 25 mission scripts, and 25 AirSim setting scripts}. 

\revmod{
\textit{Generated Scenario Blueprint Quality}: We analyzed all 25 generated scenarios to assess their relevance and their usefulness for testing sUAS. Two authors, who have both several years of experience in developing sUAS applications and testing sUAS systems, acted as evaluators independently evaluating each scenario, and provided binary ratings on (a) its relevance to the use cases and (b) its potential usefulness for testing sUAS. Both evaluators agreed that all 25 generated scenarios are relevant for the respective use case and useful for testing. This agreement resulted in a perfect Cohen's Kappa \cite{mchugh2012interrater} score  of 1.} \revmod{To expand our qualitative analysis, we used inductive coding \cite{thomas2003general} to identify recurring themes and potential gaps in the scenarios. Both evaluators first independently commented on the scenarios' unique aspects and areas for improvement, then discussed their comments to identify common themes.} 

\revmod{As part of this qualitative analysis, we found that scenarios contain \textit{Rich and Diverse Environmental Details}, such as changing wind, lighting, and temperature conditions, with realistic nuances like moderate GPS signal loss due to foliage \cite{lachapelle1994seasonal} in the area. Such subtle environmental details are often overlooked in simulation testing as they require expertise in GPS sensor signal reception and processing.} \revmod{Both evaluators also agreed that the scenarios for the Search and Rescue use cases were the most diverse ones. This finding is consistent with the diversity analysis presented in Table \ref{tab:domain-jaccard-similarity}.} 

\revmod{On the other hand, we also observed that many scenarios, while containing different mission profiles and test properties, often share a \textit{``Similar Flavor''}. For example, navigating through flooded terrain and avoiding obstacles, and  navigating through smoke-filled terrain and avoiding obstacles are two flavors of the collision avoidance test property. To avoid redundancy, human reviewers in Phase 1 must provide precise feedback to the \sagent. Our current framework allows incorporating this feedback by updating the input prompt and allowing to focus on either a broader range of scenarios or smaller nuances of more specific scenarios depending on the required testing context. In the future, this process could involve additional guidance for developers interacting with the framework, for example, by providing support for refining specific sections of the scenario blueprint by targeting and modifying selected snippets of text in \sagent's output. Our perception study with sUAS developers, discussed in Section \ref{sec:evRQ3}, further provides more insights on the usability of these scenarios in sUAS testing.}  


\revmod{\textit{Validation of Generated Executable Scripts}: Besides the qualitative analysis we further assessed the validity  (i.e., executability) of each generated script. We used \emph{Rule-based Validators} for both the \eagent and \magent to verify if the generated scripts are valid and executable using the \textit{Rules} (cf. Table \ref{tab:promptdesign}) that were enforced as part of the agents' prompt. We found that all scripts were valid and passed the validity checks, indicating that \magent and \eagent can generate valid mission scripts and simulator settings for each scenario blueprint.}
In particular, we found that the combination of \emph{Rules} and \emph{Expected Output} in the prompt ensured that the values in each JSON property accurately reflect the mission and environmental description in the scenario blueprint. These initial results indicate that 1) \sagent generates \revmod{relevant and useful} scenario blueprint, 2) the output of the \magent and \eagent can be used directly to test SuT without any human intervention, 3) the design of \sagent, \magent, and \eagent is flexible to support sUAS testing across diverse use cases, SuT, and simulation tools, and finally 4) The generation \revmod{of scripts} is fast enough to be used in large-scale automated testing. 


\begin{reqbox}{\small RQ1: \fwname Applicability \& Generalizeability}{}
\small
Based on our experiments, \fwname can be used to test two different SuT based on PX4 and ArduPilot flight controllers and also found that it can be applied to test across diverse sUAS use cases, and that valid simulation and test scripts were generated in all 25 test executions.
\end{reqbox}
\subsection{RQ2 - RAG Agents' Performance}
\label{sec:evRQ2}
We use the standard RAGAs framework \cite{es2023ragas} to quantitatively analyze the performance of the \sagent and \aagent. For the \sagent, we analyzed the 25 scenario blueprints across 5 sUAS use cases to evaluate if our knowledge base helps in generating contextually relevant and diverse blueprints. Additionally, to evaluate the \aagent's ability to produce the contextually relevant analysis report for 7 common sensor failures as well as its ability to identify issues.

\subsubsection{\textbf{\sagent RAG Performance}} 
\label{sec:rag-analysis}

 The RAG evaluation metrics scores, as presented in Table \ref{tab:s-agent-performance_metrics}, show the effectiveness of the \sagent across five use cases. Notably, the \sagent achieved high faithfulness scores (0.8-0.9), indicating that it generates factually accurate scenario blueprints and utilized real-world sUAS incident data in the knowledge-base to generate scenario blueprints. The high context precision scores (0.7-0.9) also demonstrate the agent's ability to retrieve information from the knowledge base, relevant to the sUAS use case, with minimal  accident data for a different use case. Furthermore, the high response relevancy scores (0.7-0.9) confirm that the agent generates responses relevant to the \emph{User Goals}.

However, the moderate context recall scores (0.6-0.8) suggest that the \sagent retrieves a significant portion of relevant information, but also misses some. The lower scores in \textit{Precision Agriculture} can be attributed to the relatively limited number of related incident in our knowledge-base. 

\begin{table}[h!]
\centering

\begin{tabular}{p{2.6cm}ccccc}
\toprule
\textbf{sUAS} & \textbf{Faith-} & \textbf{Context} & \textbf{Response} & \textbf{Context} \\
\textbf{Use Case}& \textbf{Fulness} & \textbf{Precision} & \textbf{Relevancy} & \textbf{Recall} \\

\midrule
Search-and-Rescue  & 0.90 & 0.80 & 0.90 & 0.80 \\
Precision Agriculture  & 0.80 & 0.70 & 0.70 & 0.60 \\
Env. Monitoring  & 0.90 & 0.80 & 0.90 & 0.80 \\
Surveillance  & 0.90 & 0.80 & 0.90 & 0.80 \\
Package Delivery  & 0.90 & 0.90 & 0.90 & 0.80 \\
\bottomrule
\end{tabular}
\caption{\sagent RAG Performance Across sUAS Use Cases
}
\label{tab:s-agent-performance_metrics}
\end{table}
\subsubsection{\textbf{Generated Scenario Blueprint Diversity}}
While the RAGA metrics offer valuable insights into the contextual relevance of the generated output, they do not evaluate the diversity of the scenario blueprints. To address this limitation, we calculated the Jaccard similarity score~\cite{bag2019efficient}, a measure of similarity between two sets ranging from 0 (completely dissimilar, which is desirable in our context) to 1 (identical), across all 5 scenarios generated for each use case. The low similarity scores, ranging between 0.24 and 0.53, presented in Table~\ref{tab:domain-jaccard-similarity}, show that the generated scenario blueprints contain a notable degree of diversity. 

\begin{table}[h!]
\centering
\begin{tabular}{p{4.9cm}r}
\toprule
\textbf{sUAS Use Case} & \textbf{Avg. Jaccard Similarity} \\
\midrule
Search-and-Rescue & 0.24 \\
Precision Agriculture & 0.41 \\
Env Monitoring & 0.44 \\
Surveillance & 0.48 \\
Package Delivery & 0.54 \\
\bottomrule
\end{tabular}
\caption{Jaccard Similarity across sUAS use cases 
}
\label{tab:domain-jaccard-similarity}
\vspace{-1em}
\end{table}

\subsubsection{\textbf{\aagent RAG Performance}}
\label{sec:agent-qualitative-analysis}
For each flight log, we generated an analysis report across 5 different test properties. For instance, test properties related to the Accelerometer included Peak Acceleration, Unexpected Acceleration Change, Average Acceleration in Each Axis, Duration of Highest Sustained Acceleration, and Consistency of Accelerometer Readings. As a result, for the seven sensor failures and flight logs, this process yielded a total of 35 responses from the \aagent, which we then analyzed using RAGAs to evaluate \aagent's performance in generating contextually relevant output. Table \ref{tab:a-agent-performance_metrics} presents the results.

Our findings indicate that the agent identified and utilized relevant parameters for each sensor to generate the analysis report. For example, in the analysis of accelerometer data, the agent selected parameters such as $accel\_bias[0]$, $failsafe$, $yawspeed$, $has\_low\_throttle$, and $accel\_fault\_detected$, and used their time-series data to construct the analysis. It is important to note that identifying relevant parameters goes beyond simple keyword matching with sensor and parameter names; the agent also utilizes the description of each parameter to identify items like $Failsafe$ and $Yawspeed$ for analyzing test properties related to the Accelerometer sensor. 
\begin{table}[h!]
\centering
\begin{tabular}{p{2.5cm}ccccc}
\toprule
\textbf{Sensor} & \textbf{Faith-} & \textbf{Context} & \textbf{Response} & \textbf{Context} \\
\textbf{Analysis}& \textbf{Fulness} & \textbf{Precision} & \textbf{Relevancy} & \textbf{Recall} \\

\midrule
Accelerometer	&	0.86	&0.88&	0.82&	0.82 \\
Gyroscope	&	0.82	&0.85&	0.82&	0.82 \\
Magnetometer	&	0.74	&0.80	&0.72&	0.75 \\
Air Speed &0.93&	0.87	&0.93&	0.83\\
Barometer &		0.80	&0.80	&0.80	&0.72 \\
Battery &0.84&	0.85&	0.87&	0.72 \\
GPS &0.90&0.86&0.90&0.80\\ 
\bottomrule
\end{tabular}
\caption{\aagent RAG Performance Across Sensor Failures}
\label{tab:a-agent-performance_metrics}
\end{table}


\subsubsection{\textbf{Generated Analysis Report Correctness}} While our previous analysis indicates that the agent can generate contextually appropriate content, assessing the accuracy of the generated report is equally important. For our dataset of 7 log files, we evaluated whether the \aagent could accurately detect the injected failures and correctly describe the actual issues in the analysis report. 

\begin{table}[htbp]
\centering
\begin{tabular}{|c |c |p{4.6cm}|r}
\hline
\textbf{Sensor Fail.} & \textbf{Issue} & \textbf{\aagent (Automated Mode)} \\
\textbf{Injected}& \textbf{Detected} & \textbf{Snippet of Analysis Report}\\

\hline

Accelerometer & \checkmark & \textit{Inconsistent accelerometer bias values} suggest issue, which can lead to inaccurate motion sensing and stability. \\ \hline

Gyroscope & \checkmark & \textit{Spikes in gyro readings} indicate potential instability. \\ \hline
Magnetometer & \checkmark & \textit{The magnetometer is occasionally experiencing faults}, This could impact the drone's stability and control. \\ \hline

Air Speed & \checkmark & the \textit{warning messages related to `airspeed off'} suggest that there were some issues with controlling the airspeed.  \\ \hline

Barometer & \checkmark & \textit{The sudden spike in altitude readings at around 150 seconds} in the baro alt meter plot could be due to a sensor error. \\ \hline

Battery & \checkmark & \textit{failsafe was activated} due to low battery capacity. \\ \hline

GPS & \checkmark &  \textit{shows sudden and erratic fluctuations} that could impact the drone's navigation and stability. \\

\hline
\end{tabular}
\caption{Sensor Issue Analysis and Agent Report 
}
\label{tab:sensor-issue-analysis}
\end{table}

We found that the \aagent correctly identified sensor failures in all 7 flight logs as shown in Table \ref{tab:sensor-issue-analysis}. The generated report provides clear textual descriptions of data indicating a failure based on \aagent's interpretation of time-series plots. For instance, the agent was precisely able to find an issue with the barometer sensor by identifying a sudden or unusual spike in the $baro\_alt\_meter$ reading at second 150 of the flight. These results highlight the effectiveness of \aagent in automatically analyzing simulation logs and supporting developers in analysis. 


\begin{reqbox}{\small RQ2: \fwname Performance}{}
\small
 The \sagent generated relevant and diverse scenarios for each sUAS use case. The \aagent generated contextually relevant analysis and automatically identify common sensor failures.
\end{reqbox}
 
\subsection{RQ3 - End-Users perception of \fwname}
\label{sec:evRQ3}

\begin{table}[]
    \centering
    \begin{tabular}{|L{1cm}|p{2.4cm}|p{4cm}|}
    \hline
         \textbf{P.\#}& \textbf{sUAS Dev Exp.}  & \textbf{Flight Log Analysis Experience}\\\hline
        P1& 1 Year & 10+ flight logs \\ \hline
        P2& 5 Years & 50+ flight logs \\ \hline
        P3& 3 Years & 10+ flight logs \\ \hline
        P4& 7 Years & 10+ flight logs  \\ \hline
    \end{tabular}
    \caption{Participants' Details}
    \label{tab:participants}
    \vspace{-1.5em}
\end{table}
We recruited 4 sUAS developers for interviews from our network. \revmod{Details} of each participant is shown in Table \ref{tab:participants}. \revmod{All participants interacted with \aagent, while three participants interacted with \sagent to evaluate and provide feedback on the perceived usefulness of the framework. P4 could not engage with \sagent due to logistical issues.}

\subsubsection{\textbf{Problem Validity}} \revmod{P1 validated the inherent difficulty in manually generating thousands of diverse scenarios and highlighted the value of providing automation support to this process.} P2, P3, and P4 confirmed that analyzing hundreds of flight log parameters is a time-consuming task and heavily depends on the developer's experience. According to P2, \textit{``these drones output hundreds of time-series data, and when something goes wrong, analyzing that is a real challenge"}. The participants unanimously agreed that automation support would greatly simplify simulation testing for them. 
\revmod{\subsubsection{\textbf{Perceived Usefulness of Generated Scenarios}} P1, P2, and P3 highlighted that generated scenario blueprints are contextual as per their input prompt and include details such as \emph{high winds}, \emph{temperature}, \emph{gps quality}, and \emph{locations}. P3 specifically mentioned: ``the details in the generated scenario are pretty impressive such as flight duration of 30 minutes and payload of 1.5 kg''. While our study participants agreed that the generated scenarios are highly detailed and read like realistic situations, they also mentioned that current state-of-the-art sUAS simulation tools do not support direct execution of such complex scenarios. They emphasized that executing these scenarios would require significant technical effort and substantial enhancements to the simulation tools.
}
\revmod{\subsubsection{\textbf{Perceived Usability of \aagent}} All participants appreciated the inclusion of plots for relevant parameters extracted from flight logs in the reports and mentioned \emph{significant time-saver}(to analyze thousands of parameters). They highlighted that the automated identification and plotting of relevant time-series data is particularly valuable, as it addresses a common challenge: determining where to begin their analysis. For example, P2, an expert in flight log analysis, entered the prompt: \emph{``Was the satellite count low?''} In response, the agent correctly plotted data for parameters such as \emph{satellite\_used} and additionally reported a supporting parameter, \emph{rx\_message\_lost\_count}, which P2 found interesting. Based on this interaction with \aagent, P2 remarked: \emph{``the analytics agent provided really good analysis.''}}. 


\subsubsection{\textbf{Improvement Opportunities}} \revmod{When interacting with \sagent, P2 recommended that the agent should generate ``\textit{incremental test scenarios}''. For instance, the first scenario could involve a simple flight path without obstacles, while subsequent scenarios could gradually increase complexity by adding obstacles, limited battery availability, and varying weather conditions. This structured progression would help developers build a comprehensive test suite.} 

\revmod{On the other hand, when interacting with \aagent,} P2, suggested incorporating knowledge from public forums (e.g., \url{https://discuss.ardupilot.org}) into \aagent's knowledge base. This would enable \aagent to also propose potential fixes to issues. P1 and P3 recommended displaying additional information on the User Interface, such as flight paths, to facilitate targeted queries about specific flight duration or timestamps. Finally, P4 stated that the entire framework could be best taken advantage of during a CI/CD pipeline to get an analysis report after every commit to SuT.

\begin{reqbox}{ \small RQ3: sUAS Developers' Perception of \fwname}{}
\small
 sUAS Developers expressed enthusiasm for utilizing the framework, and indicated that \fwname would serve as a valuable tool during their simulation analysis activities.

\end{reqbox}
\section{Discussion and Lessons Learned}
\label{sec:discussion}

\subsubsection{\textbf{Application to CPS Testing}}The proposed framework's multi-agent design allows for broad application to various CPS, including Autonomous Ground Vehicles (AGV) such as self-driving cars and delivery robots. Adapting the agents' Knowledge Base and Prompt Design is key to applying the \fwname to AGVs.For instance, to test the navigation capabilities of an autonomous car in rainy weather, the \sagent can generate a scenario blueprint using instances of road accidents stored in its knowledge base. This blueprint can be utilized by the \magent to produce a list of destinations, specifying the locations the car must visit during simulation. For widely used simulation tools such as CARLA~\cite{malik2023carla}, the \eagent can automatically generate the necessary configuration files (e.g., \textit{CarlaSettings.ini}) to initialize the CARLA simulation environment \cite{malik2023carla} with rainy conditions. ultimately, the generated simulation logs can be analyzed by the \aagent, leveraging its comprehensive domain knowledge of analyzing parameters of autopilots of autonomous cars. Therefore, the concept of \fwname is not limited to sUAS and has the potential to be applied for more general CPS testing.

\subsubsection{\textbf{Realism of sUAS Simulations}} While our \fwname was successful in generating scenario blueprints and executable scripts, we observed a huge gap between the scenario blueprint we generated and the realism of actual simulation execution. \revmod{For instance, a scenario blueprint generated for river search-and-rescue use case included simulating the movement pattern of a drowning person in the river. However, simulating such dynamism 
is not supported by Airim or similar sUAS simulation tools.} 
This was also emphasized by participant P3. Therefore, our study also highlights the need for more advanced sUAS simulation tools that allow simulating context specific elements of the environment.


\subsubsection{\textbf{Swarm sUAS Analysis}} \aagent is a valuable tool in detecting sensor failures and potential end-users, both novice and experts, perceive this as a useful tool in their simulation analysis process. However, real-world flights suffer from failures that extend beyond sensors, particularly those arising from the human interaction with sUAS during the flight \cite{vierhauser2021hazard}. Additionally, state-of-the-art flight analytics tools are limited to supporting single sUAS analysis, which falls short for swarm sUAS applications that require simultaneous analysis of multiple sUAS. Therefore, we will expand the scope of \aagent in future to analyze variety of sUAS failures and support swarm sUAS analysis.

\section{Threats to Validity}
\label{sec:threats}
\begin{itemize}[leftmargin=*]
    \item Our feasibility and generalizability tests indicate the broad applicability of the framework. However, these tests are based on assumptions about inputs to more complex SuTs. To mitigate this limitation, we are working to collaborate with our industry partners to apply our framework to more advanced and complex SuTs in future work.

    \item 
    \revmod{The standard approach to compute relevance and faithfulness metrics for evaluating the RAG performance of \sagent and \aagent relies on another LLM, introducing a construct validity threat. To address this, we will construct a database of testing scenarios across major sUAS use cases that serve as representative benchmarks. These scenarios will act as ground truth, allowing us to compute scores such as BLEU~\cite{papineni2002bleu} to fairly compare the quality of LLM-generated scenarios with human-generated ones.}
    
    \item Our perception study offers preliminary insights into the framework's usefulness from the perspective of sUAS experts. However, a formal user study is needed to obtain deeper insights into the framework's effectiveness compared to existing sUAS simulation testing and analysis practices.

\end{itemize}


\section{Related Work}
\label{sec:relwork}
Testing CPS is a rather broad research area, ranging from assessing the accuracy of onboard AI models~\cite{chandrasekaran2021combinatorial} applying fuzzing and metamorphic testing for generating test cases~\cite{wang2021robot}. 
In this section, we focus on (1) \revmod{sUAS simulation tools and testing,} (2) the use of LLMs for testing, and (3)Automated Analysis of sUAS simulation results.




\revmod{ \bulletitem{sUAS Simulation Testing:} In the domain of sUAS, researchers have developed various different simulation tools that support both Software-in-the-Loop (SITL) and Hardware-in-the-Loop (HITL) testing \cite{giovagnola2023airloop} such as Gazebo~\cite{gazebo}, AirSim \cite{shah2018airsim, bondi2018airsim}, and RFlySim \cite{dai2021rflysim}. These platforms serve as the foundation for testing their performance under controlled conditions. Besides these core simulation capabilities, the research community has further enhanced and adapted these platforms to meet specific sUAS testing needs, including LiDAR sensor simulation for testing sensor-based algorithms, environmental conditions like wind for evaluating sUAS navigational capabilities \cite{zhang2023dronereqvalidator,dronewis,agrawal2023requirements}. 
In this context, our framework can also be leveraged to generate initialization/configuration files, for example, specifying LiDAR sensor sensitivity or noise levels, and wind configurations. Furthermore, while these tools provide valuable support to developers and researchers, our framework offers an end-to-end simulation testing solution that covers scenario specification, execution, and analysis.}


\bulletitem{Testing using LLMs:}
\revmod{Recently, various ML-based approaches have emerged, particularly LLMs, for testing.} For example, Arora~\etal~\cite{arora2024generating} used LLMs for test scenario generation. While their approach performs well and tests various functional aspects, it lacks a concrete structure to automatically execute the scenarios. For autonomous vehicles, Chang~\etal~\cite{chang2024llmscenario} propose LLMScenario, using prompt engineering for scenario generation, while Zhang~\etal~\cite{zhang2024chatscene} employ an LLM with Scenic Programming Language for creating safety-critical CARLA scenarios, though limited by its reliance on a specific language. In contrast, \fwname is a language-agnostic, end-to-end framework for sUAS simulation testing, enabling seamless scenario generation, execution, and analytics across various simulation tools. \revmod{Further, we used a RAG-based approach to decouple our framework from the underlying LLM. This provides flexibility to switch between state-of-the-art or objective-specific models such as Trustworthiness \cite{hua2024trustagent} or Functional Safety \cite{shi2024aegis} to capture diverse perspectives when generating scenario blueprints.}

\bulletitem{Automated sUAS Flight Analytics:}
Recent studies have demonstrated the efficacy of deep learning techniques in anomaly detection for automated analysis of sUAS flights\cite{al2022resam,islam2024adam, ma2023detecting}. By identifying unexpected spikes in sensor data, these methods provide valuable insights into abnormal sUAS behavior. Traditional anomaly detection approaches typically employ a bottom-up analysis, where expert developers leverage their prior knowledge to perform an in-depth examination of specific sensor data and diagnose the root cause of sUAS abnormal behavior. However, this approach has a significant limitation: it requires extensive domain-specific knowledge and expertise to analyze the complex sensor data. In contrast, our proposed \aagent employs a top-down approach, where developers initiate analysis with high-level investigation questions, identifying relevant flight controller parameters and drilling down to determine underlying causes of issues.
\section{Conclusion}
\label{sec:conclusion}
In this paper, we have presented a novel multi-LLM-agents-based framework, automating the sUAS simulation testing process. Our extensive experiments have demonstrated the framework's capabilities in testing diverse SuT, generating contextually relevant and diverse scenario blueprints, effectively analyzing flight logs, and providing valuable support to both novice and experienced developers in analyzing and understanding simulation results.
The implications of \fwname are twofold: (a) enabling exhaustive simulation testing and rapid iteration between development and simulation tests, and (b) facilitating the development and safe deployment of sUAS by substantially reducing the time and effort required for testing and validation, while achieving high simulation test coverage. For future work, we have planned additional studies with industry partners to validate the usability of \fwname in testing more advanced and complex SuT. Furthermore, we are planning to extend our \aagent to support automated analysis of swarm sUAS.

\section{Data Availability}
All supplementary materials, including the framework codebase, are  available at \url{https://github.com/UAVLab-SLU/AutoSimTestFramework}.

\section{Acknowledgment}
The work in this paper was funded under USA National Aeronautics
and Space Administration (NASA) Grant Number: 80NSSC23M0058

\bibliographystyle{abbrv}
\bibliography{icse2025,droneusecases}

\end{document}